\documentclass[aps,twocolumn]{revtex4-1}

\usepackage{graphicx}

\begin{document}
\title{Tri-critical point and suppression of the Shastry-Sutherland phase in Ce$_{2}$Pd$_{2}$Sn by Ni doping}

\author{J.G. Sereni$^{1}$, G. Schmerber$^2$, M. G\'omez Berisso$^{1}$, B.
Chevalier $^3$ and J.P. Kappler$^2$}
\affiliation{$^1$ Div. Bajas Temperaturas, CAB-CNEA and Conicet, 8400 S. C. de Bariloche, Argentina\\
$^2$ IPCMS, UMR 7504 CNRS-ULP, 23 rue de Loess, B.P. 43 Strasbourg Cedex 2, France\\
$^3$ CNRS, Universit\'e de Bordeaux, ICMCB, 87 av. Dr. Schweitzer,
33608 Pessac Cedex, France}

\date{\today}

\begin{abstract}

{Structural, magnetization and heat capacity measurements were
performed on Ce$_2$(Pd$_{1-x}$Ni$_x$)$_2$Sn ($0 \leq x \leq 0.25$)
alloys, covering the full range of the Mo$_2$FeB$_2$ structure
stability. In this system, the two transitions observed in
Ce$_2$Pd$_2$Sn (at $T_N=4.8$\,K and $T_C=2.1$\,K respectively)
converge into a tri-critical point at $T_{cr}\approx 3.4$\,K for
$x\approx 0.3$, where the intermediate antiferromagnetic AF phase
is suppressed. The $T_N(x)$ phase boundary decrease is due to an
incipient Kondo screening of the Ce-4f moments and local atomic
disorder in the alloy. Both mechanisms affect the formation of
Ce-magnetic dimers on which the Shastry-Sutherland lattice (SSL)
builds up. On the contrary, the $T_C(x)$ transition to the
ferromagnetic ground state increases as a consequence of the
weakening of the AF-SSL phase. Applied magnetic field also
suppresses the AF phase like in the stoichiometric compound.}

\vspace{0.5cm} Keywords: Cerium compounds, Critical Points,
Magnetic Phase Diagrams, Magnetic transitions

 $^*$ E-mail-address of corresponding author:
jsereni@cab.cnea.gov.ar

\end{abstract}


 \maketitle

\section{Introduction}

Crystalline structures with local symmetries which favor magnetic
frustration attract special interest because they provide the
scenario for novel phases formation. In many cases, magnetic
frustration or exotic phases occur in competition with classical
long range order ground states GS. Thus, the search of exotic
phases is addressed to the vicinity of magnetic transitions, since
the 'roughness' of the free energy may develop new relative minima
as a function of a non trivial order parameter \cite{Kirchpat}.
Those minima compete in energy for the formation of novel phases,
which may become unstable under small variation of control
parameters like magnetic field, alloying or pressure.

Beside the mentioned frustrated states, alternative phases may
occur under peculiar geometrical conditions. Among them, the so
called Shastry-Sutherland lattice (SSL) \cite{Shastry,Miyahara}
builds up from mutually orthogonal magnetic dimers which impose
further topological and magnetic constrains. These conditions are
realized in some members of the R$_{2}$T$_{2}$X family of
compounds (with R = Rare Earths or Actinides, T = Transitions
Metals and X = 'p' type metalloids), which crystallize in
tetragonal Mo$_{2}$FeB$_{2}$ structure \cite{Peron93}. In that
structure, each 'R' layer forms a mosaic of magnetic atoms
coordinated as isosceles triangles between nearest- and next
nearest neighbors centered on the $z$-axis of the T element. The
shortest side of those triangles, where R-R magnetic dimers form,
is shared by two consecutive triangles like in the shortest
diagonal of a rhombohedron. The resulting simple square-lattice of
mutually orthogonal rhombohedron mimicks a sort of 'pinwheel'
centered on the $z$-axis of the X element \cite{Ce2Pd2SnNofield},
whereas the net of dimers form a simple two dimensional (2D)
square lattice.

Recently, SSL phases were found in a number of R$_{2}$T$_{2}$X
compounds \cite{Kim,Ce2Pd2SnNofield} and the involved magnetic
interactions theoretically discussed \cite{Lacroix}. In
Ce$_{2}$Pd$_{2}$X compounds, where Ce-Ce dimers form due to a
nearest neighbor ferromagnetic FM interaction, the SSL phase shows
up within a limited range of temperature (c.f. between $T_N=4.8K$
and $T_C=2.1K$ in Ce$_{2}$Pd$_{2}$Sn \cite{Ce2Pd2SnNofield}).
Below $T_C$, a FM-GS takes over undergoing a first order
transition due to the discontinuity in the order parameter.
Further studies performed under magnetic field on the mentioned
compound \cite{Ce2Pd2Sn_field} showed that the intermediate phase
is suppressed applying a magnetic field $B_{cr}\approx 0.12$\,T at
$T_{cr} \approx 3.2$\,K. The fact that the SSL phase is suppressed
by quite low magnetic field in Ce$_{2}$Pd$_{2}$Sn remarks its
instability respect to a 3D FM magnetic structure as GS.

\begin{figure}
\begin{center}
\includegraphics[angle=0,width= 0.5 \textwidth] {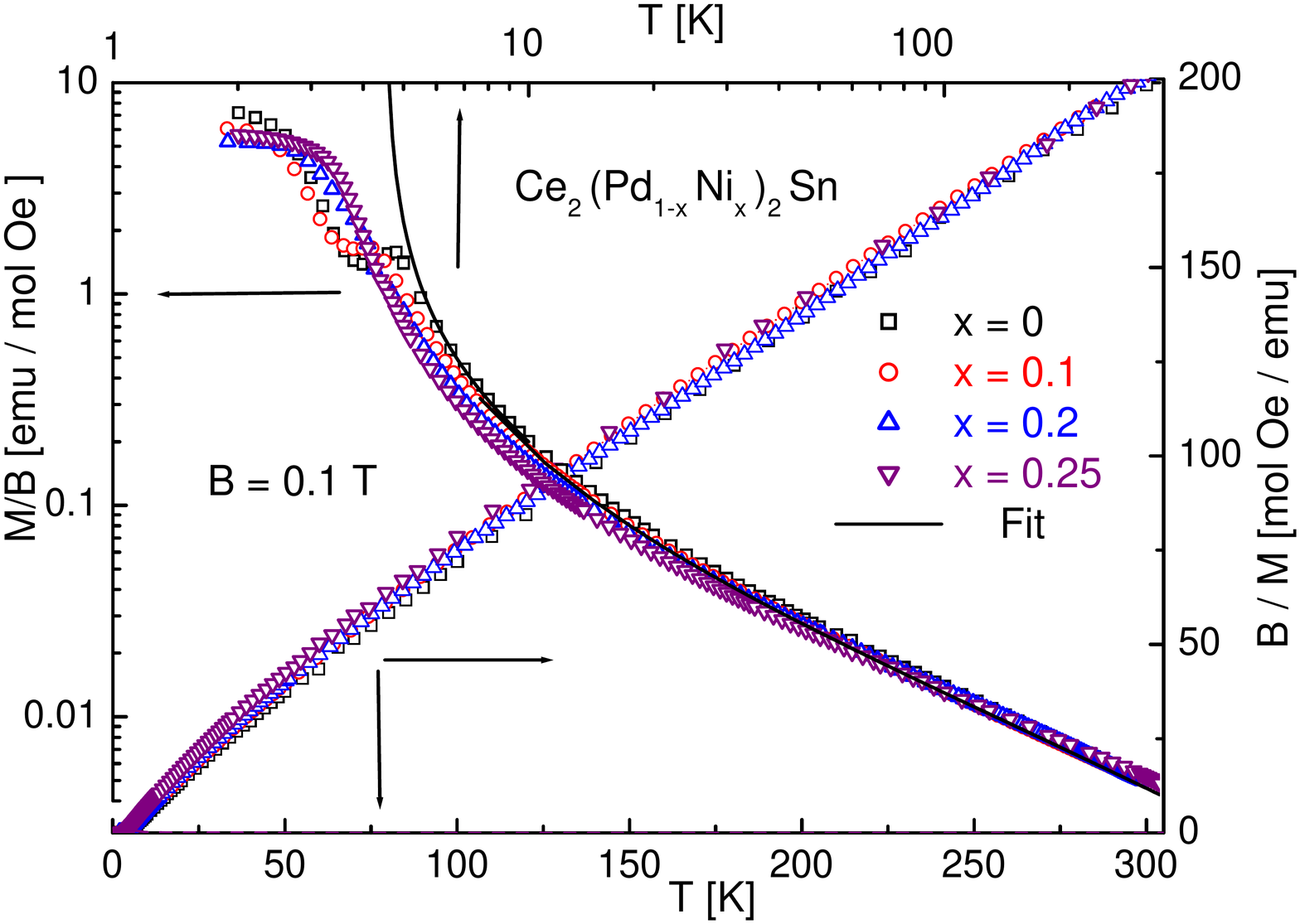}
\end{center}
\caption{(Color online) High temperature magnetic susceptibility
in a Log-Log representation for different Ni concentrations (left
and upper axes) and inverse susceptibility (right and lower
axes).} \label{F1}
\end{figure}

In this work we have investigated Ni doped
Ce$_{2}$(Pd$_{1-x}$Ni$_{x}$)$_{2}$Sn alloys with the scope to
compare the effect of structural pressure and magnetic field on
the stability of the SSL phase. Since Pd and Ni are iso-electronic
elements, but being Ni atoms about $25\%$ smaller in volume than
Pd ones, an effective structural pressure is expected to weak the
Ce-4f magnetic moments. Similarly, applied magnetic field shall
progressively suppress the SSL phase like in the stoichiometric
compound.

\begin{figure}
\begin{center}
\includegraphics[angle=0,width= 0.5 \textwidth] {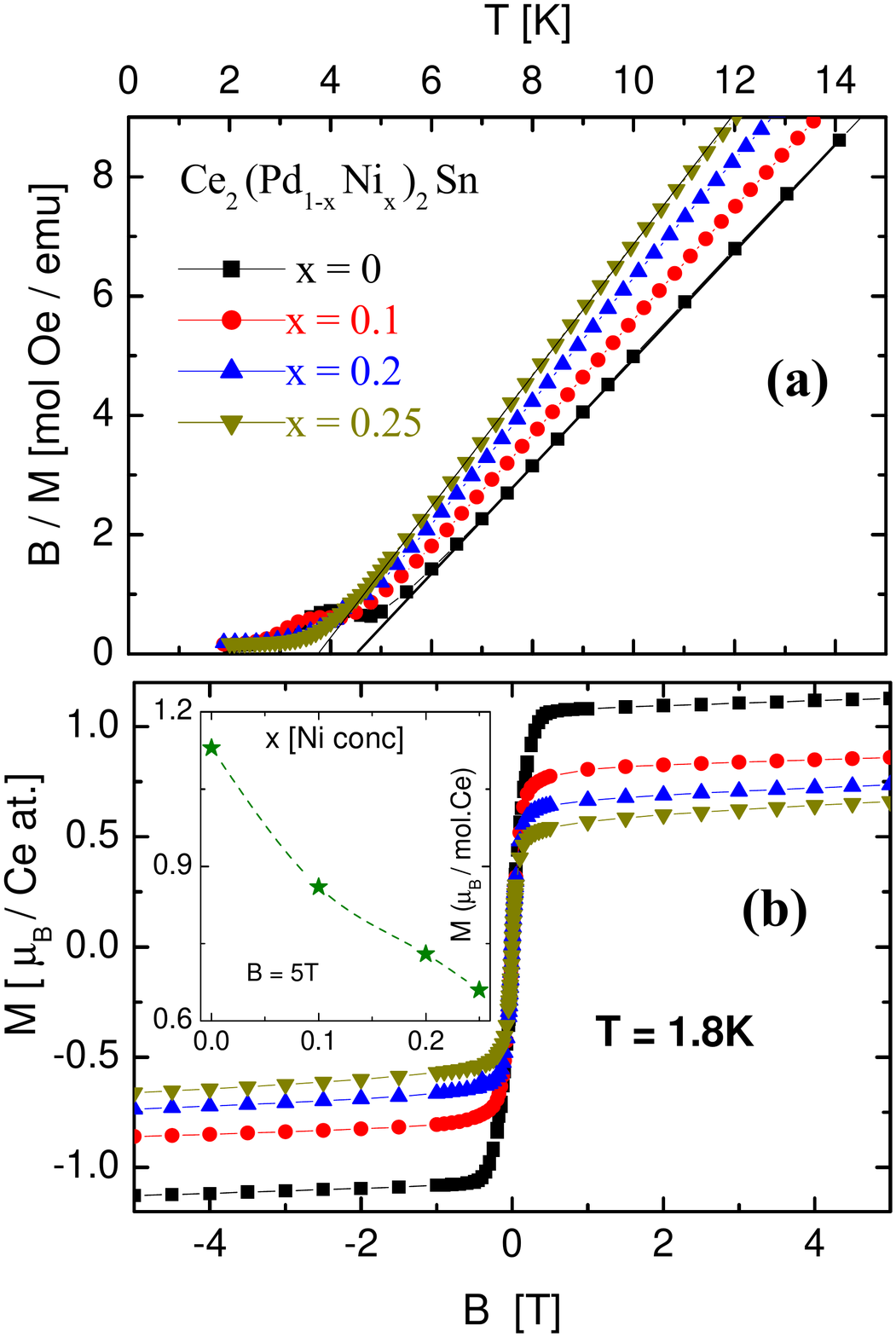}
\end{center}
\caption{(Color online) (a) Low temperature inverse magnetic
susceptibility showing the variation of $\theta_P^{LT}$, measured
with $B=0.1$\,T. (b) Field magnetization within the FM phase,
showing the decrease of the saturation moment with Ni
concentration increase. Inset: magnetic moment at $B = 5$\,T
within the $x\leq 0.25$ range.} \label{F2}
\end{figure}

In a former study performed on this system \cite{CeNiIn}, no
change of structure was reported under Ni doping. The unit cell
volume was reported to follow a Vergard's law up to $x=0.6$, with
the paramagnetic Curie-Weiss temperature $\theta_P$ decreasing
from 20\,K at $x=0$ to -40\,K at $x = 0.5$. Electrical resistivity
($\rho$) was observed to be nearly temperature independent with a
high value of $\rho_0$ and large difference upon cooling and
heating, which was interpreted as due to cracks in the alloyed
samples \cite{CeNiIn}. However, in a recent investigation, a
change of structure was detected at $x=0.35\pm 0.05$
\cite{CePdNi2SnArxiv}. Thus, the large $\rho_0$ values and the
abnormal $\rho(T)$ dependence can be explained as due to non
single crystalline phases in the concentration region where the
system presents a coexistence of two crystalline structures.

\section{EXPERIMENTAL DETAILS AND RESULTS}

Details for sample preparation were described in a previous paper
\cite{Ce2Pd2SnNofield}. Structural characterization confirms the
single phase composition of the samples in a tetragonal
Mo$_2$FeB$_2$-type structure for $x<0.3$. Beyond a short range of
coexistence of two phases, this system stabilizes in an
orthorhombic W$_{2}$CoB$_{2}$ type structure for $x>0.4$
\cite{CePdNi2SnArxiv}. The actual composition of the
stoichiometric compound was determined to be
Ce$_{2.005}$Pd$_{1.988}$Sn$_{0.997}$ after SEM/EDAX analysis.
Lattice parameters slightly decrease with Ni concentration from
$a=7.765\AA$ and $c=3.902\AA$ at $x=0$, to $a=7,7122\AA$ and
$c=3.8941\AA$ at $x = 0.25$. These variations drive a reduction of
the unit cell volume of about $\approx 1.5\%$ between $x=0$ and
0.25. On the contrary, the 'c/a' ratio practically does not
change, indicating that the local symmetry is not affected by
doping.

\begin{figure}
\begin{center}
\includegraphics[angle=0,width= 0.6 \textwidth] {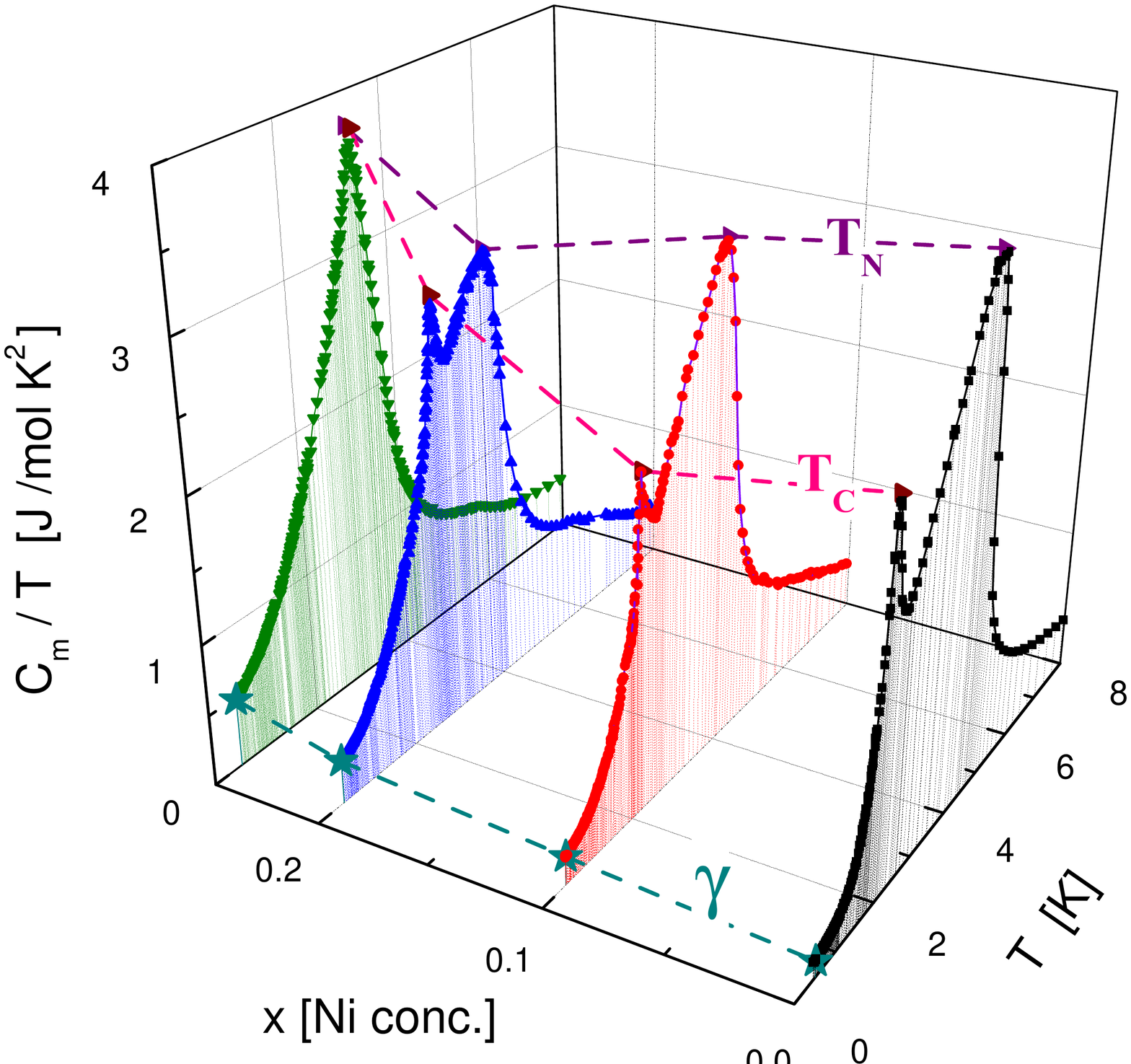}
\end{center}
\caption{(Color online) Evolution of the magnetic specific heat
divided $T$ for Ce$_2$(Pd$_{1-x}$Ni$_x$)$_2$Sn showing the
convergent variation of both transitions AF-$T_N$ and FM-$T_C$
converging at $x_{cr}\geq 0.25$. Dashed line labeled $\gamma$
represents a rising heavy fermion contribution.} \label{F3}
\end{figure}

DC-magnetization measurements were carried out using a standard
SQUID magnetometer operating between 2 and 300\,K, and as a
function of field up to 5T. Specific heat was measured using
standard heat pulse technique in a semi-adiabatic He-3 calorimeter
in the range between 0.5 and 20\,K, at zero and applied magnetic
field up to 4T. The magnetic contribution to the the specific heat
$C_m$ was computed subtracting the phonon contribution extracted
from the isotypic La$_2$Pd$_2$Sn compound. Electrical resistivity
was measured between 0.5\,K and room temperature using a standard
four probe technique with an LR700 bridge. However, sample
cracking inhibits to extract valuable information from $\rho(T)$
data.

\subsection{Ni concentration dependent properties}

As shown in Fig.~\ref{F1}, Ni doping does not affect significantly
the magnetic susceptibility ($\chi$) at high temperature. Fits of
$\chi(T)$ do not show significant variation of crystal field
effect (CEF) from the values observed in stoichiometric
Ce$_2$Pd$_2$Sn \cite{Ce2Pd2SnNofield}, with respective splitting
at $\Delta_I=65\pm5$\,K and $\Delta_{II}=240\pm10$\,K. From the
inverse of $\chi(T)$, one extracts that the high temperature
$\theta_p^{HT}$ slightly increases only for $x\geq 0.2$ from
-16\,K to -25\,K. This indicates that only a minor hybridization
increase occurs for the CEF excited levels within this range of Ni
doping. Coincidentally, the high temperature Ce magnetic moment
practically does not change from its Ce$^{3+}$ value.

A different behavior is observed at low temperature, where
$\theta_p^{LT} > 0$. Its value decreases from 4.5\,K (at x = 0) to
3.7\,K (at x = 0.25), as shown in Fig.~\ref{F2}a, indicating a
moderate Ni doping effect on the doublet GS magnetism. This goes
hand by hand with the magnetic saturation extracted from field
dependent magnetization measurements (see Fig.~\ref{F2}b),
performed at $T=1.8$\,K within the FM phase. Also the cusp of
$M(T)$ at $T=T_N$ is reduced till it is overcome by the FM
contribution (see Fig.~\ref{F4}a).

In the case of specific heat, Ni concentration affects quite
significantly the $C_m(T)$ jumps of both transitions as shown in
Fig.~\ref{F3}. While the temperature of the upper one decreases
from $T_N(x=0)=4.8$\,K to $T_N(x=0.25)\approx 3.5$\,K, the lower
transition increases from $T_C(x=0)=2.1$\,K up to
$T_C(x=0.25)\approx 3.3$\,K (see inset in Fig.~\ref{F4}a). These
opposite Ni dependencies drive the transitions to merge at a
tri-critical point at $T_{cr}\approx 3.4$\,K, just above the
highest Ni concentration studied ($x=0.25$).

$T_N(x)$ is a second order transition, whereas $T_C(x)$ is of
first order. This character is recognized by an hysteretic
$C_m(T)$ dependence measured around $T_C$ and reflects the
discontinuity in the magnetic order parameter at $T=T_C$. Both
$C_m(T)$ maxima decrease with Ni doping, however as they merge
into an unique maximum the $C_m(T)$ cusp clearly rises. Notice the
sharpness of the cusp in sample $x = 0.25$ revealing its vicinity
to the critical point.

\subsection{Magnetic field effects}

\begin{figure}
\begin{center}
\includegraphics[angle=0,width= 0.5 \textwidth] {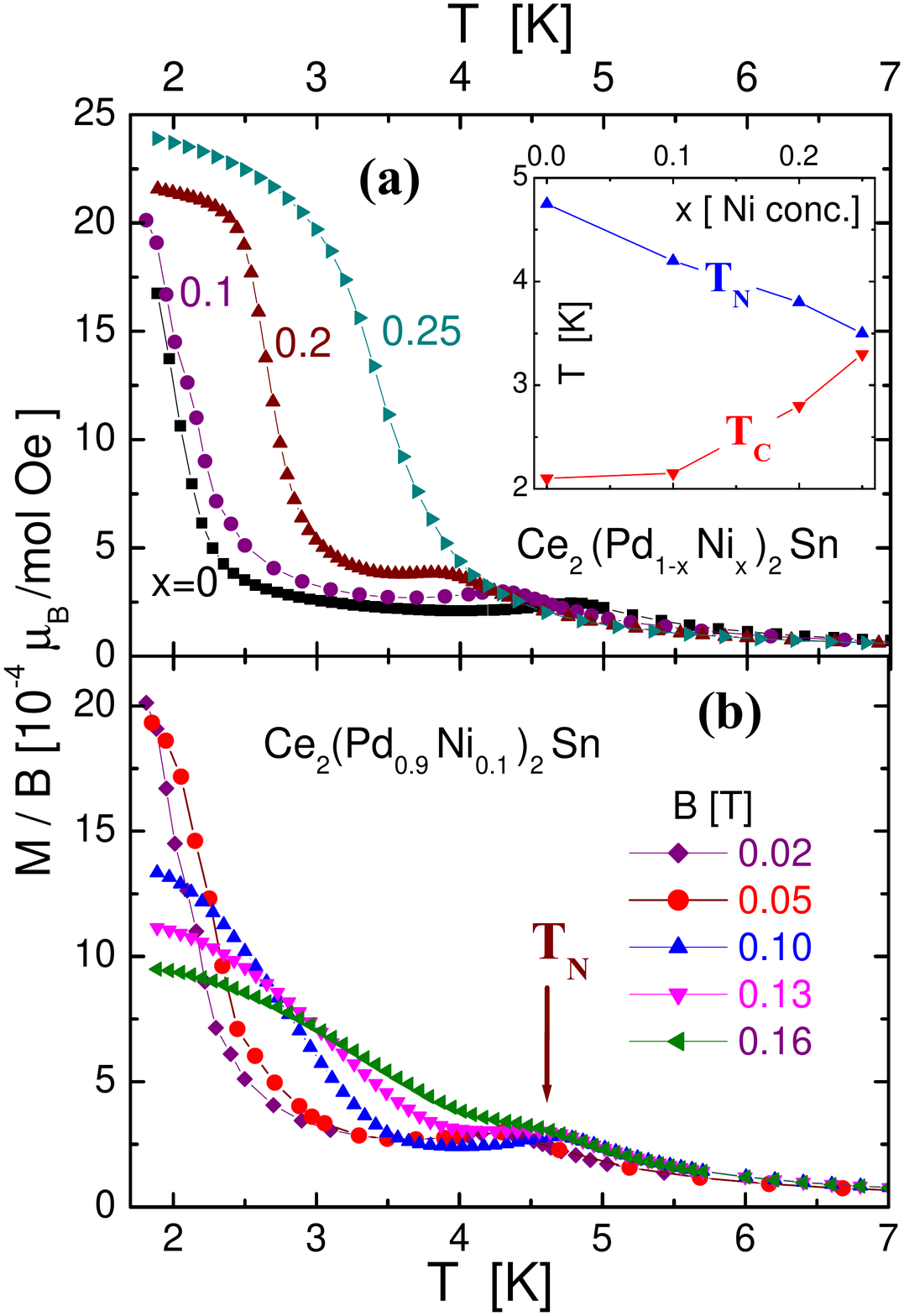}
\end{center}
\caption{(Color online) Comparison of the thermal dependence of
magnetization around AF and FM transitions: (a) as a function of
Ni doping (with $B=0.02$\,T). Inset: $T_N(x)$ and $T_C(x)$
dependencies. (b) Magnetic field effect on the $x=0.1$ sample.}
\label{F4}
\end{figure}

Doping and magnetic field effects coincide enhancing the FM
character of the GS phase with the consequent $T_C(x,B)$ increase.
On the contrary, while Ni doping produces a decrease of $T_N(x)$
magnetic field practically does not affect the upper transition.
The doping effect can be observed in Fig.~\ref{F4}a, which shows
how the temperature of the spontaneous magnetization (related to
$T_C$) increases, overlaping the $M(T_N)$ cusp as Ni content
increases.

In the case of applied magnetic field, one can observe in
Fig.~\ref{F4}b for sample $x=0.1$ how the temperature of the
$M(T_N)$ cusp remains practically unchanged. The same behavior was
observed in the stoichiometric compound \cite{Ce2Pd2Sn_field} and
is confirmed in sample $x=0.2$. In the case of sample $x=0.25$,
the $T_N$ cusp is already overlaped by the FM signal at very low
applied field. One have to remind that, apart from the $T_C(x)$
increase, the FM transition is smeared under applied magnetic
filed and therefore the high temperature tail of the FM
magnetization overcomes the $T_N$ cusp faster than the actual
increase of $T_C(x)$.

\section{DISCUSSION}

\subsection{Specific heat Gap}

\begin{figure}
\begin{center}
\includegraphics[angle=0,width= 0.5 \textwidth] {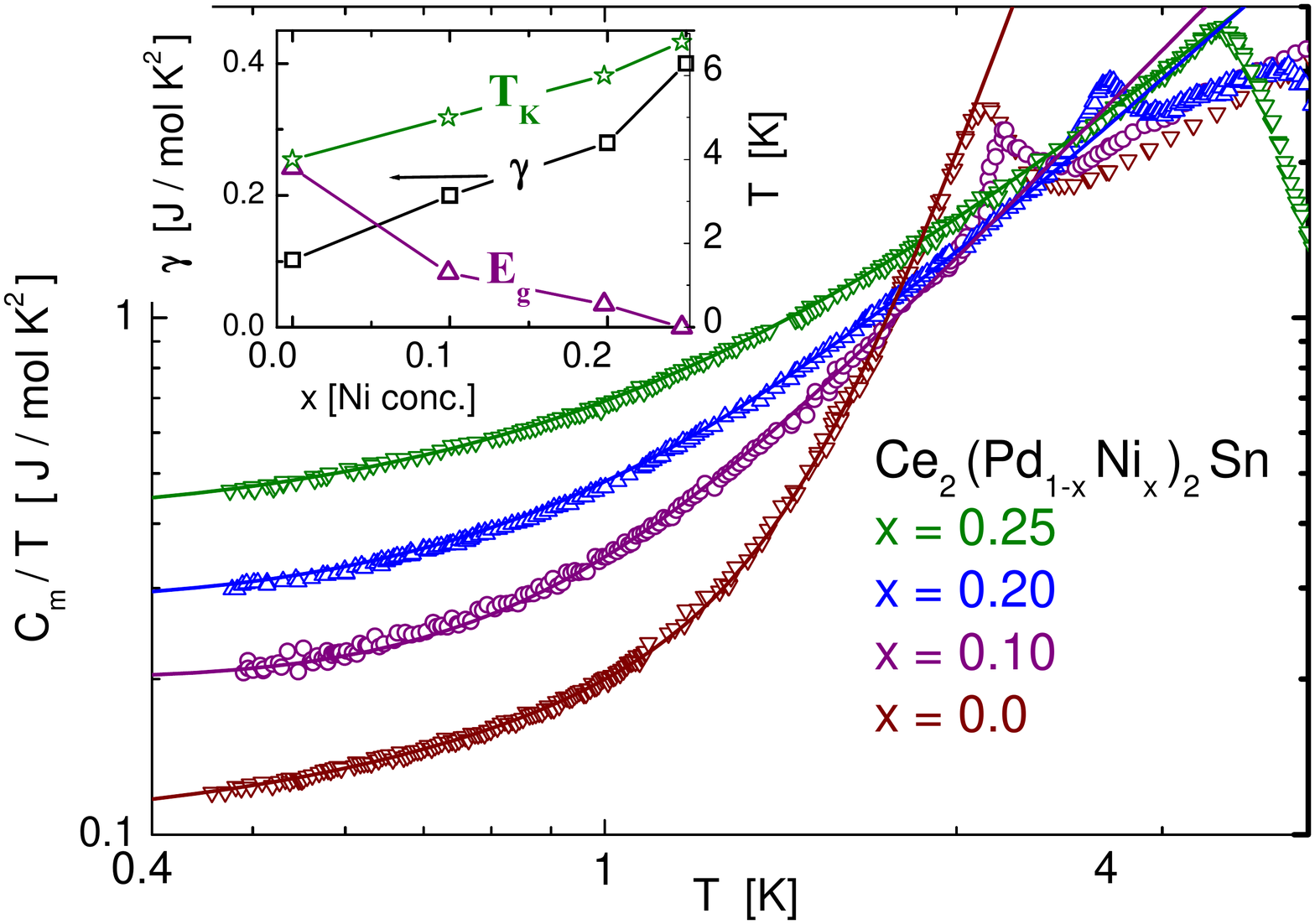}
\end{center}
\caption{(Color online) Analysis of the $C_m(T)/T$ dependence for
$T<T_C$ in a Log-Log representation. Continuous curves are fits
from which $\gamma$ and magnon gaps $E_g$ are extracted, see the
text. Insert: Ni concentration dependence of $\gamma(x)$ (left
axis), and $E_g(x)$ and Kondo temperature $T_K(x)$ (right axis).}
\label{F5}
\end{figure}

Relevant information related to the magnetic structure of this
system can be extracted from the analysis of the thermal
dependence of the specific heat within the ordered phase. Since
the $x=0$ sample have shown strong anisotropic effects, we have
analyzed the magnetic contribution of the FM phase to $C_m(T)$ by
fitting all curves with an unique function in order to evaluate
the variation of the characteristic parameters as a function of Ni
content. The simplest applicable function is $C_m/T = \gamma + A
T^2 exp(-E_g/T)$, where the $\gamma$ term accounts for the degrees
of freedom behaving as a Kondo liquid and $E_g$ represents the
energy (expressed in temperature) of the gap of anisotropy in the
magnon spectrum. Respective fits are included in Fig.~\ref{F5}
using a double logarithmic scale, and computed values are
collected in the inset.  These results indicate that the specific
heat behaves as that of systems whose gap of anisotropy deceases
to zero when approaching the critical concentration.
Coincidentally, the $\gamma$ contribution increases with Ni
content.

\subsection{Entropy and Kondo Temperature}

\begin{figure}
\begin{center}
\includegraphics[angle=0,width= 0.55 \textwidth] {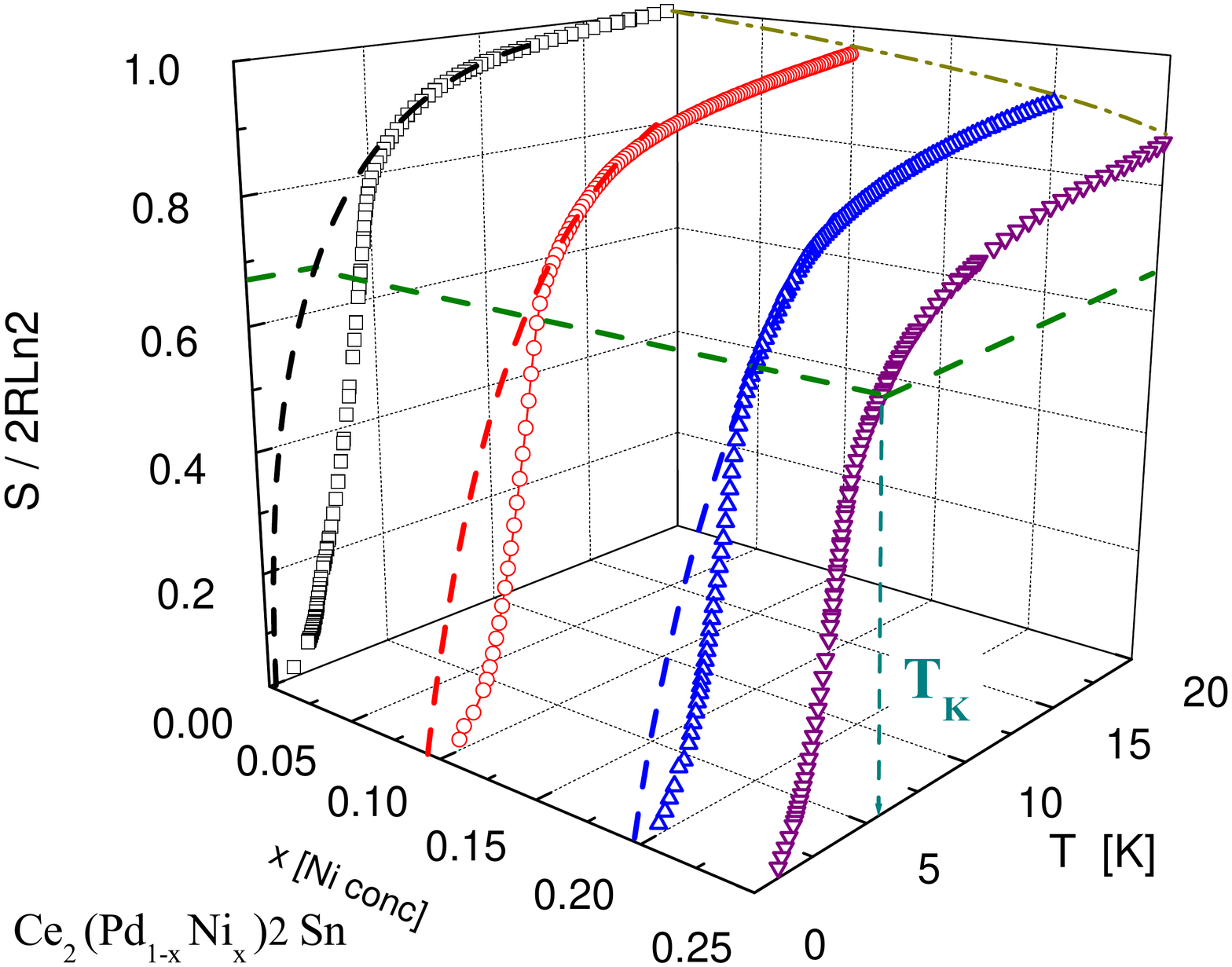}
\end{center}
\caption{(Color online) Thermal dependence of the entropy for the
studied samples in a 3D representation. Dashed curves indicate the
extrapolation of $S_m(T)$ from $T>T_N$ to $T<T_N$, and the
straight lines the $S_m=0.66 2RLn2$ value from which $T_K$ is
extracted.} \label{F6}
\end{figure}

From the thermal variation of the magnetic contribution to the
entropy ($S_m$), the effect of Ni doping on the Ce-4f magnetic
moment can be traced. As it can be observed in Fig.~\ref{F6}, it
is only for $x\geq 0.2$ that some incipient hybridization effect
can be detected. To extract the Kondo temperature variation, we
have applied the current Desgranges-Schotte criterium \cite{Desgr}
from which at $T=T_K$ the entropy reaches the value $S_m=0.66
RLn2$. Notice that in our case the comparison has to be done in
$2RLn2$ units because there are two Ce atoms contained in a
formula unit of this compound. Since this model was proposed for
single impurities, the $S_m(T)$ curve has to be extrapolated from
$T>T_N$ following the curvature within the non-interacting
(paramagnetic) phase, which does not account for the condensation
of degrees of freedom into the ordered state. Despite of the small
error introduced by this extrapolation, the relevant information
extracted is the low value of $T_K$ ($T_K\approx 4$\,K for $x=0$)
and its variation with concentration ($T_K \approx 7$\,K for
$x=0.25$). The computed $T_K(x)$ values are included in the inset
of Fig.~\ref{F5}. There is an apparent contradiction in this
coincident increase of $T_K(x)$ and $\gamma(x)$ because they are
expected to depend inversely to each other (i.e. $\gamma \propto
T_K$). This behavior can be explained by the fact that $T_K$ is an
{\it intensive} parameter (i.e. a scale of energy) which increases
with $x$, and $\gamma$ is {\it pseudo-intensive} because it
depends on the degrees of freedom involved in the Kondo liquid
contribution. From this comparison we conclude that there is a
gradual transference of degrees of freedom from the ordered phase
to the Kondo liquid component.

\subsection{Stability of the SSL phase versus doping}

\begin{figure}
\begin{center}
\includegraphics[angle=0,width= 0.5 \textwidth] {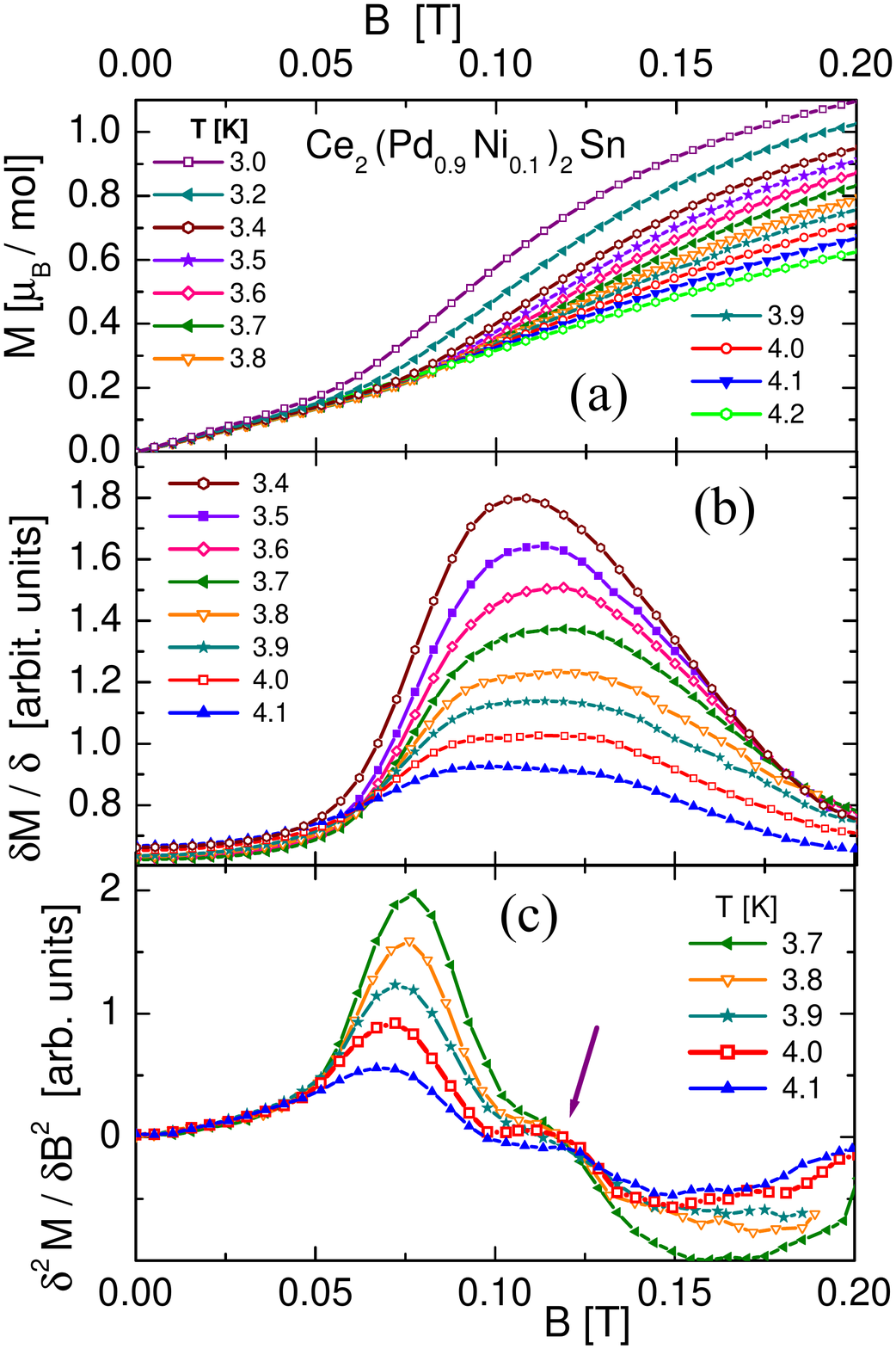}
\end{center}
\caption{(Color online) (a) Field dependence of magnetization of
sample $x=0.1$ between 3 and 4.2\,K. (b) First derivative and (c)
second derivative for the $3.7\ leq T \leq 4.1$\,K range, with the
arrow indicating the satellite anomaly.} \label{F7}
\end{figure}

In order to investigate up to which extend Ni doping affects the
stability of the SSL phase, we have studied the magnetic field
effect in the alloyed samples following the same procedure like in
the stoichiometric compound ($x=0$) \cite{Ce2Pd2Sn_field}.
Starting with the $x=0.1$ sample, we have performed $M(B)$
measurements within the range of temperature where the features
related to the SSL phase could be detected, see Fig.~\ref{F7}a.
Above $T_C$, the $M(B)$ isotherms show a 'S' shape reflecting an
AF character, which is rapidly polarized by magnetic field. There
is however a crossing of magnetization curves around B=0.11T
followed by a slight modulation which was taken as indication of a
SSL formation in the stoichiometric compound
\cite{Ce2Pd2Sn_field}. Further information was extracted from the
$\partial M/\partial B|_T$ derivative shown in Fig.~\ref{F7}b. An
incipient shoulder can be appreciated at $B\approx 0.12$\,T in the
$3.7\leq T \leq 4.1$\,K isotherms. Since this feature was taken as
an indication of the incipient $M(B)$ step related to the SSL
phase, we have proceeded to analyze also the second $\partial ^2
M/\partial B ^2|_T$ derivative of the original $M(B)$ results. The
respective curves are collected in Fig.~\ref{F7}c, providing
better information to determine the onset of the induced FM phase
(main maximum at $B\approx 0.077$\,T) and the satellite anomaly
(see the arrow at $B\approx 0.12$\,T.

\begin{figure}
\begin{center}
\includegraphics[angle=0,width= 0.55 \textwidth] {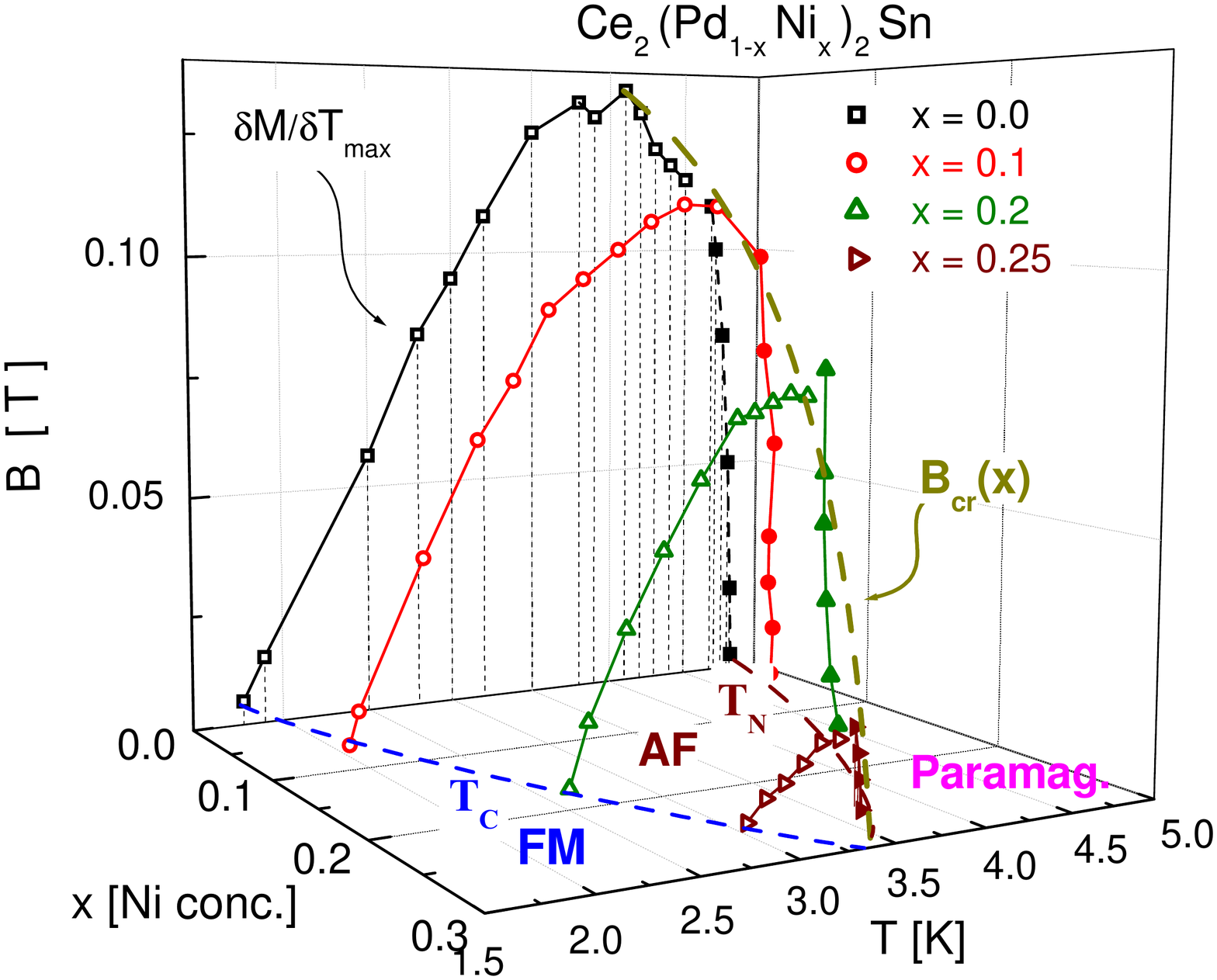}
\end{center}
\caption{(Color online) Ni concentration and magnetic field
dependent phase diagram in a 3D representation.} \label{F8}
\end{figure}

In comparison with the stoichiometric compound, the main maximum
has decreased from $B\approx 0.11$ to $0.08$\,T, whereas the
satellite shifted from $B\approx 0.15$ to $0.12$\,T with a
significant reduction in the intensity. Although the $B$ values
follow the predictions for a SSL for the appearance of a plateaux
in the magnetization (at 1/4 and 1/8 of the saturated moment
\cite{Miyahara}), these results from sample $x=0.1$ are in the
limit of our experimental detection. Effectively, in the following
concentration studied ($x=0.2$) no traces of the mentioned anomaly
were observed indicating a border line for the SSL formation at
$x\simeq 0.1$. Not only the weakening of Ce magnetic moments but
also the local disorder due to the difference between Pd and Ni
sizes inhibit the Ce-Ce dimers formation with the consequent
smearing of the SSL phase.

\subsection{Magnetic phase diagram}

The comparison between doping and magnetic field effects is
presented in a 3D phase diagram in Fig.\ref{F8}. There, the field
driven transition between intermediate AF and FM-GS phases is
drawn according to the temperature of the maximum slope of $M(T)$
(i.e. $\partial M/\partial T\mid_{max}$) which is in agreement
with $C_m(T)$ results. Both phase boundaries join at a critical
point, which decreases with Ni content from $T_{cr}(x=0)=(4.2\pm
0.3$)\,K and $B_{cr}=(0.16\pm 0.02$)T to $T_{cr}(x=0.25)\approx
(3.5\pm 0.3$)\,K and $B_{cr}\approx (0.02\pm 0.005)$\,T, at the
edge of the intermediate phase disappearance (i.e. where
$B_{cr}\to 0$). Unfortunately, the tri-critical point where AF, FM
and paramagnetic phases join together cannot be reached by doping
because it lies beyond the limit of stability of the Mo$_2$FeB$_2$
crystal structure.

The evolution of AF and FM phase boundaries respond to different
effects. Starting from high temperature, the decrease of the upper
(AF) transition $T_N(x)$ is expected from i) the competition
between Kondo effect and RKKY magnetic interactions according to
Doniach-Lavagna model \cite{Doniach} acting on the Ce-4f magnetic
moments intensity and ii) the local atomic disorder introduced by
alloying atoms of different sizes. Both effects weaken the
formation of magnetic Ce-dimers on which the SSL builds up. On the
contrary the $T_C(x)$ increase is a consequence of the
intermediate AF phase weakening like in Ce$_2$Pd$_2$Sn under
applied magnetic field \cite{Ce2Pd2Sn_field}.

The effect of magnetic field is qualitatively similar for each
concentration accounting that the range of stability of the AF
phase is progressively reduced. Notably, the $\partial M/\partial
B\mid_T$ derivatives of sample $x=0.25$ still display a weak
maximum in a restricted range of field and temperature. This
indicates that the critical Ni concentration is not reached yet
despite in the $C_m(T)$ cusp of that sample both phase boundaries
seem to have converged. Interestingly, the critical temperature
$T_{Cr}$ reached by doping at zero field and by magnetic field
applied on the stoichiometric compounds are very similar. This
confirms the equivalent effect of both parameters on the
suppression of the intermediate phase.

\section{Conclusions}

We have seen that Ni doping on Ce$_2$Pd$_2$Sn strongly affects the
intermediate AF phase of this compound. While the $T_N(x)$
transition decreases, $T_C(x)$ increases following a first order
transition line with an end critical point. In fact, at zero
field, $T_N(x)$ and $T_C(x)$ converge right above the highest Ni
concentration studied ($x=0.25$). The increase of the FM-GS phase
is simply due to the fact that the 2D-AF character weakens respect
to the 3D-FM magnetic interactions ones the inter-plan interaction
becomes relevant. The progressive weakening of the magnon gap to
its disappearance coincides with this behavior.

The SSL phenomenology was hardly recognized from the $M(B)$
dependence only in the $10\%$ doped Ni sample, for higher
concentrations (i.e. $x=0.2$) none of those symptoms have been
recognized. Since hybridization effects are only marginal, one
concludes that atomic disorder plays an important role inhibiting
the SSL phase formation. Magnetic field further weakens the
intermediate phase like in the stoichiometric compound, and allows
to better define the tri-critical point extrapolated at
$T_{cr}\approx 3.4$\,K and $x=0.3$, where $B_{cr} \to 0$.
Unfortunately $x=0.3$ lies within the structural instability
range. Nevertheless, further attempts to get closer to that
critical point, and to investigate the Ni-rich side of the phase
diagram are in progress.

\section*{Acknowledgments}
The author acknowledges L. Amigo and J. Luzuriaga for magnetic
measurements. This work was partially supported by PICTP-2007-0812
and SeCyT-UNCuyo 06/C326 projects.

\end{document}